\documentclass{IEEEcsmag}

\usepackage[colorlinks,urlcolor=blue,linkcolor=blue,citecolor=blue]{hyperref}

\usepackage{upmath}

\jvol{XX}
\jnum{XX}
\paper{-}
\jmonth{February}
\jname{arXiv Preprint}
\pubyear{2023}

\usepackage[dvipsnames,svgnames]{xcolor}
\definecolor{mred}{rgb}{.80,.12,.30}
\definecolor{MRED}{rgb}{.80,.12,.30}
\definecolor{grey}{rgb}{0.5,0.5,0.5}
\definecolor{lgrey}{rgb}{0.7,0.7,0.7}
\definecolor{purple}{rgb}{.75,0,.85}
\definecolor{pistachio}{rgb}{0.58, 0.77, 0.45}
\definecolor{traingreen}{HTML}{4CAF50}
\definecolor{myorange}{rgb}{0.94, 0.36, 0.13}
\definecolor{limegreen}{rgb}{0.2, 0.8, 0.2}
\definecolor{brightlavender}{rgb}{0.75, 0.58, 0.89}

\usepackage{extarrows}
\usepackage{ccicons}

\newif\ifnotes
\notestrue

\begin{document}

\sptitle{arXiv Preprint}

\title{Visualizing Uncertainty in Sets}

\author{Christian Tominski}
\affil{University of Rostock, DE}

\author{Michael Behrisch}
\affil{Utrecht University, NL}

\author{Susanne Bleisch}
\affil{FHNW University of Applied Sciences and Arts Northwestern Switzerland, CH}

\author{Sara Irina Fabrikant}
\affil{University of Zurich, CH}

\author{Eva Mayr}
\affil{University for Continuing Education Krems, AT}

\author{Silvia Miksch}
\affil{TU Wien, AT}

\author{Helen Purchase}
\affil{Monash University, AU}

\markboth{Visualizing Uncertainty in Sets}{Visualizing Uncertainty in Sets}

\begin{abstract}
\looseness-1
Set visualization facilitates the exploration and analysis of set-type data. However, how sets should be visualized when the data is uncertain is still an open research challenge. To address the problem of depicting uncertainty in set visualization, we ask (i) which aspects of set type data can be affected by uncertainty and (ii) which characteristics of uncertainty influence the visualization design. We answer these research questions by first developing a conceptual framework that brings together (i) the information that is primarily relevant in sets (i.e., set membership, set attributes, and element attributes) and (ii) different plausible categories of (un)certainty (i.e., certainty, undefined uncertainty as a binary fact, and defined uncertainty as quantifiable measure). Based on the conceptual framework, we systematically discuss visualization examples of integrating uncertainty in set visualizations. We draw on existing knowledge about general uncertainty visualization and fill gaps where set-specific aspects have not yet been considered sufficiently.
\end{abstract}

\maketitle

\chapteri{A}nalysing set data encompasses consideration of the sets themselves, the elements within the sets, and attributes of both the sets and the elements. Take for example academic courses at a university (e.g., Biology, Mathematics) as sets, and the students enrolling in those courses as the set elements. Set visualizations aim to express such set-type data visually to support analysis and better understanding. Relevant set analytical questions involve set memberships (i.e., who is enrolled in which course), set cardinality (i.e., how many students are in a course), or set intersections (i.e., which course combinations are favored by students). Such kinds of set analytical questions can be answered, for example, with the help of Euler diagrams, Venn diagrams, and bipartite node-link representations. While set visualizations themselves are an active research frontier~\cite{Alsallakh16SetVisSTAR}, there are far fewer research activities focusing on the implications of uncertainty for set visualization~\cite{Bleisch23DagstuhlReport, JenaEDRP20, BonneauHJOPRS14}. For our courses-and-students example, we might not know exactly how many students are enrolled in a course or how old they are.

In fact, it is challenging to design visual representations of sets where uncertainty is involved. This is because both the set data themselves and also the information about their uncertainty need to be communicated to a reader. The interpretation of this uncertainty has a major impact on decisions that are made based on the data, not only for simple applications such as course planning, but also for more complex scenarios like comparison of ensemble forecasting models or gene-to-phenotype mapping.

So far, the literature offers little insight into the implications of uncertainty for set visualization~\cite{Bleisch23DagstuhlReport}. In particular, a distinction of classes of uncertainty in the context of set-type data is missing. Only if we know, however, what types of uncertainty are relevant for set-type data can we design expressive visual representations. Therefore, the main objective of this paper is to systematize uncertainty considerations for set visualization. We propose a conceptual framework that brings together (a) different facets of set data that might be affected by uncertainty, and (b) different types of uncertainty that might influence the visualization design. As the primarily relevant data facets in sets, our framework contains: set membership, set attributes, and element attributes. In terms of different types of (un)certainty, we distinguish: certainty, undefined uncertainty as a binary fact, and defined uncertainty as quantifiable measure.

From this proposed framework, we derive interesting combinations of data facets and types of uncertainty that are calling for dedicated visualization strategies. While some cases can be addressed with existing visualization approaches, others seem to be more intricate to deal with and lack suitable solutions. We discuss exemplary cases with the goal to identify concrete gaps in the literature and to sketch initial thoughts on how these gaps can be closed. 

Before developing our framework, we will next introduce basic set and uncertainty terminology.

\section{Sets \& Uncertainty}

Set theory has been investigated in mathematical logic in the nineteenth century by Cantor~\cite{Cantor1895Sets} to describe collections of objects, called sets, and their elements. Sets do not impose any ordering on their elements. Sets may overlap, making well-defined relations between sets possible, including containment, exclusion, and intersection. Moreover, both sets and elements may have various attributes associated with them. Accordingly, the primarily relevant data characteristics~($D$) for set-type data are: (i) set membership, (ii) set attributes, and (iii) element attributes.

Uncertainty ($U$) relates to information that is unknown, vague, or of varying accuracy. So, a good starting point is to think about what is known and what is unknown. In a perfect world, we know the data and assume they are accurate. There is no uncertainty, which we denote as $U=0$. For set-type data this means that we know for certain the sets and the elements, their membership, and their attributes. 

However, in the real world, uncertainty is commonly encountered in everyday life~\cite{Lindley13Uncertainty}. It is inherent to any piece of information and thus also present in any dataset, data model, or visualization and has been studied in many scientific disciplines and academic fields~\cite{gershon1998visualization, MacEachren2005, MacEachren12UncertaintyVisualization, Sacha2016, Pelz2021}. Given the universal relevance of uncertainty, it is not surprising to find various notations and categorizations in different fields. Terms like aleatoric uncertainty or stochastic uncertainty are relevant in mechanical engineering~\cite{Pelz2021}. But also more common terms such as incertitude, probability, or ignorance appear in the context of uncertainty.

The common theme behind the heterogeneous landscape of terminology is that uncertainty is present in all parts of the data-driven scientific research process~\cite{NCGIA1998}, starting with measurement and data capture, data transformations and processing, data modeling and visualization, and finally human inference and decision making with visual data displays~\cite{Sacha2016, Mason2016}. Uncertainty in data sources can be of locational, temporal, and/or semantic nature~\cite{NCGIA1998}. Uncertainty issues in data capturing can stem from, e.g., data provenance, acquisition methods, and measurement inaccuracies. Uncertainty also arises and gets further propagated in data transformations and processing, including data modeling. Uncertainty can also occur in data portrayal methods, and lead to perceptual issues of the viewer of uncertainty-depicting visualizations. Finally, uncertainty might arise in human interpretations and decision making. As a result, we argue that uncertainty should always be also considered in sets and their visualization. 

An important question to ask is how much do we actually know about the uncertainty in our set-type data? Here, we distinguish two scenarios. One scenario is that we know that there is uncertainty, but we cannot tell accurately where it is, what it is, or how much of it exists. In other words, we know for a fact that uncertainty is present in our data, but no further details. We denote this as $U > 0$. In the second scenario, we also know that uncertainty exists, and we know with certainty where, what, and how much of it is in our data. For the sake of simplicity, we denote this as $U = p$. The letter $p$ is a strong simplification of what could be known about the uncertainty and $p$ can take different forms. When set membership of an element $a$ and a set $X$ is certain, one can say either $a \in X$ or $a \notin X$. Under uncertainty, $p$ might denote a probability of $a$ being a member of $X$, $P(a, X) = p$, which is a notation known from fuzzy sets. We could also say that $p$ denotes a more complex probability distribution (e.g., $p = \mathcal{N}(\mu,\,\sigma^{2})$) based on which set membership is decided. In relation to the data attributes of elements or sets, we may understand $p$ as the probability value or probability distribution of an attribute taking a particular data value. Additionally, it is common for uncertain attribute values to specify them via a range of possible values, in which case $p = [l, u]$ is some interval with a lower and upper bound of $l$ and $u$.

Overall, the characteristics of set data $D$ and the types of uncertainty $U$ form the basis for a conceptual framework of uncertainty in set visualization, which will be described next.

\begin{table*}
	\caption{Framework of uncertainty in set visualization with relevant set characteristics and categories of (un)certainty. \ccby}
	\label{tab:framework}
	\centerline{\includegraphics[width=\textwidth]{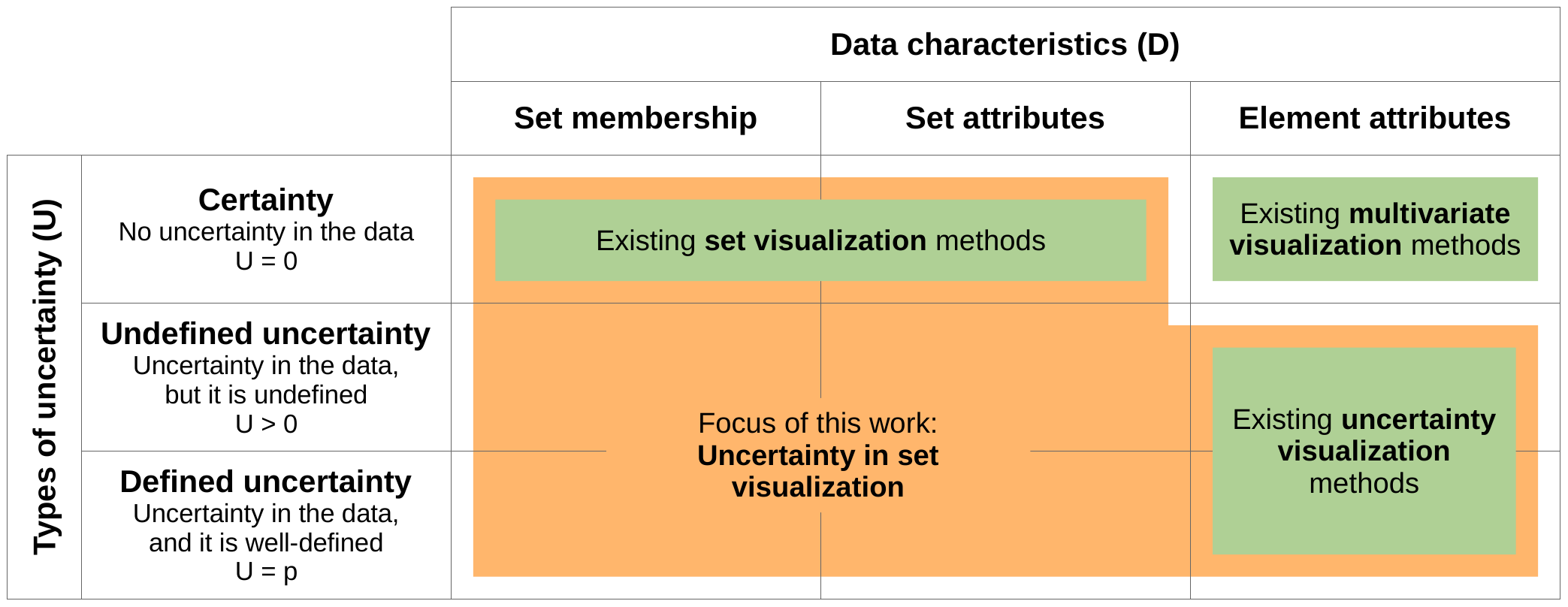}}
\end{table*}

\section{A Framework for Uncertainty in Set Visualization}

In terms of data characteristics $D$, the framework distinguishes: \textbf{set membership}, \textbf{set attributes}, and \textbf{element attributes}. Related to uncertainty $U$, we use the different plausible types of (un)certainty: \textbf{certainty} ($U = 0$), \textbf{undefined uncertainty} as a binary fact ($U > 0$), and \textbf{defined uncertainty} as quantifiable measure ($U = p$). We captured the framework in \textbf{Table~\ref{tab:framework}}, whose columns and rows respectively represent $D$ and $U$. The cells of the table correspond to different combinations of data characteristic and type of uncertainty for which adequate visualization methods are needed.

The most interesting cells in Table~\ref{tab:framework} are marked in orange, and will be described later. For the green cells in the table, one can resort to or draw inspiration from established visualization methods. Three areas are relevant in this context. First, when the data are certain ($U=0$), multivariate visualization methods can be used to depict element attributes. Second, for certain set memberships and set attributes, one can use existing set visualization methods. Third, when the attributes of data elements are uncertain ($U>0$ or $U=p$), we are moving into the area of uncertainty visualization.

For the first area, multivariate visualization, we refer to the existing visualization literature~\cite{Ward15DataVis, Tominski20IVDA}. For the second and third areas, we provide some more details below as they can inform the design of uncertain set visualizations.

\subsection{Set Visualization}

\begin{figure}
	\centerline{\includegraphics[width=\columnwidth]{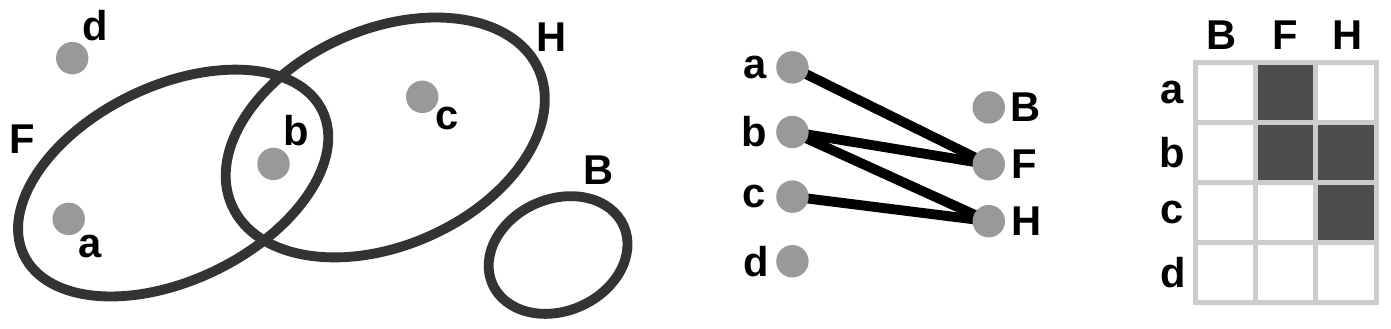}}
	\footnotesize\hspace{7em}(a)\hspace{9em}(b)\hspace{6em}(c)
	\caption{Common visual representations of sets include Euler/Venn diagrams (a), bipartite node-link diagrams (b), and matrices (c), all representing the same data. \ccby}
	\label{fig:setvis}
\end{figure}

The survey by Alsallakh et al.~\cite{Alsallakh16SetVisSTAR} provides a comprehensive overview of general set visualization methods. Common examples for visualizing fundamental set data are Euler/Venn diagrams, bipartite node-line diagrams and matrix visualizations (see \textbf{Figure~\ref{fig:setvis}}). Other visualization methods are based on overlays, aggregation, and scatterplots.

Different set visualization techniques support different set-analytic tasks, which can be structured according to the data values and data dimensions they are tackling~\cite{Alsallakh16SetVisSTAR}. Examples are tasks related to elements (find, count, filter), sets (cardinality, count), set relations (intersection, union), and element attributes (determining the highest attribute value in set), as well as combinations thereof, such as finding/selecting elements that belong to a specific set, deriving the number of sets in a set family, analyzing intersection relations, e.g., find out if a certain pair of sets overlap, or if a certain group of sets overlap, i.e., have a non-empty intersection, or find out the attribute values of a certain element.

\begin{figure*}
	\centerline{
		\includegraphics[width=.24\textwidth]{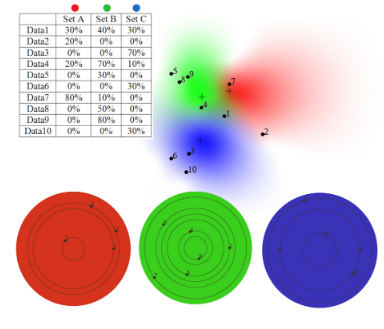}\hfill
		\includegraphics[width=.24\textwidth]{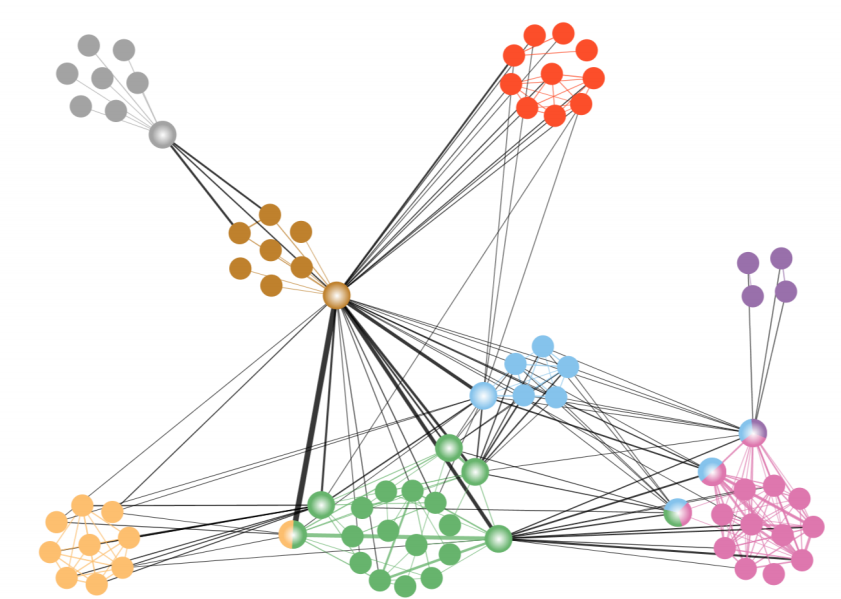}\hfill
		\includegraphics[width=.24\textwidth]{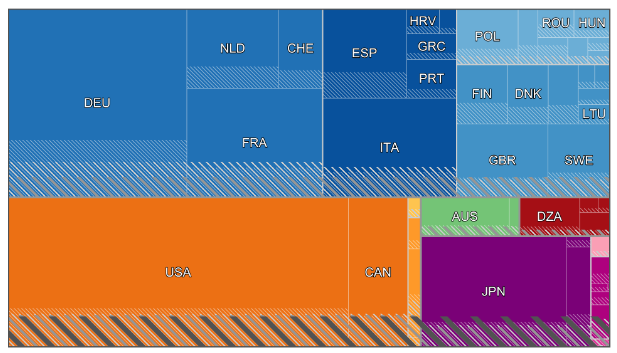}
		\includegraphics[width=.24\textwidth]{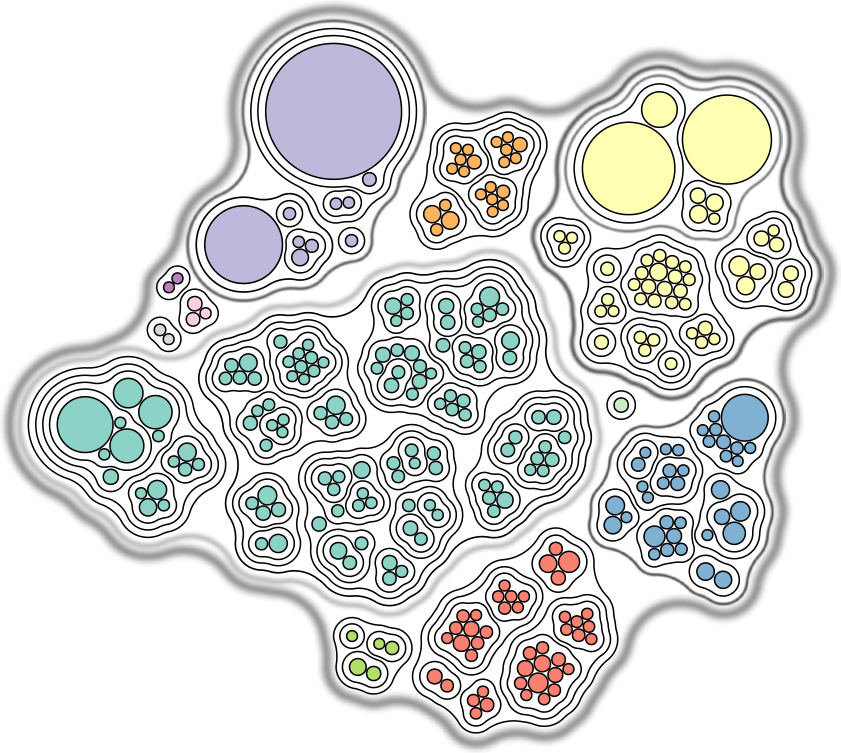}\hfill
	}
	\footnotesize\hspace{7em}(a)\hspace{13em}(b)\hspace{13em}(c)\hspace{13em}(d)
	\caption{Prior work on visualizing defined uncertainty and set membership in (a) Venn diagrams~\cite{Zhu2018} and (b) node-link diagrams~\cite{Vehlow2013} as well as visualizing uncertainty and set size in (c) treemaps~\cite{Sondag2020} and (d) with circle packing~\cite{Goertler2018}.
	}
	\label{fig:setvisuncertainty}
\end{figure*}

\subsection{Uncertainty Visualization}

Designing visual representations of uncertain data is challenging, mainly due to the fact that not only the data $D$ themselves need to be encoded visually, but also the information about their uncertainty $U$ needs to be communicated. Above all, visualization users must be able to extract all the encoded information (both the data and their uncertainty) from the visualization, which can be formulated abstractly as a pipeline, inspired by algebraic visualization design~\cite{Kindlmann14Algebraic}:

\[
(D, U) \xlongrightarrow{~~m~~} V \xlongrightarrow{~~i~~} (D', U').
\]

The visualization designer defines a mapping $m$ of data $D$ and uncertainty $U$ to create a visual representation $V$. Through an interpretation $i$ of the visual representation $V$, human observers extract their own versions of data $D'$ and uncertainty information $U'$. The scientific challenge, in this auto-encoder-like pipeline, is to understand the cognitive process of $i$ and to devise mappings $m$ so that ideally $D = D'$ and $U = U'$ for all human observers. The congruence of $D$ and $D'$, as well as $U$ and $U'$, can serve as a guiding principle for the visualization of uncertain data.

There are many ways to depict data uncertainty~\cite{gershon1998visualization, griethe:2006:UDMP,Brodlie2012,Mason2016,BonneauHJOPRS14, Duebel173DGeoUncertainty, JenaEDRP20}. Much empirically-grounded research on the representation of uncertainty exists in geospatial visualization \cite{MacEachren2005,MacEachren12UncertaintyVisualization,Kinkeldey2014, McKenzie2016,Robinson2019}, graph visualization \cite{Guo2015}, statistical visualization \cite{Hullman2015}, and temporal visualization \cite{Gschwandtner2016,Kay2016,Bors20SegmentationUncertainty}.

The empirically validated uncertainty visualization framework proposed by MacEachren and colleagues~\cite{MacEachren2005, MacEachren12UncertaintyVisualization} is an attractive candidate to directly transfer more broadly to the depiction of uncertainty. MacEachren et al. first empirically assessed the intuitiveness of a visual variable (e.g., location of symbol, size, color value, color hue, color saturation, texture, orientation, etc.) to judge the suitability of abstract or iconic point symbols for depicting data variation in a given category of uncertainty. Second, they also measured the relative performance of the most intuitive point symbol depiction of uncertainty with a focus on symbol effectiveness for a typical use task: assessing and comparing the aggregate uncertainty in two regions of a graphic display. Their stimuli are generic enough so that findings can be transferred to many data visualization types and use cases, including the visualization of set elements. 

Based on their studies, MacEachren et al. proposed generalizable design guidelines including the visual variables fuzziness and relative location and distance (from a known location in the center of a cross-hairs) to work particularly well for the depiction of uncertainty in point symbols. Color value and arrangement are also rated highly by their study participants. Both size and transparency are potentially usable. Color saturation of the point symbol, often cited as intuitively related to uncertainty, was ranked quite low in their study. Later we apply MacEachren et al.'s uncertainty visualization framework to the case of visualizing uncertain element attribute values. 

\medskip

Now that we have dealt with the 'easier' green boxes from Table~\ref{tab:framework}, we will move on to discuss the 'tougher' orange box, the visualization of uncertain set characteristics.

\subsection{Uncertainty in Set Visualization}

 In comparison to general uncertainty visualization, the representation of uncertain set data has received less attention. A few attempts exist (see \textbf{Figure~\ref{fig:setvisuncertainty}}) for the visualization of fuzzy sets \cite{Vehlow2013, Zhu2018}, where set membership has a defined uncertainty. For set attributes, we are aware of only two prior works, which represent set size with a defined uncertainty in set hierarchies using a treemap~\cite{Sondag2020} and a circle packing visualization \cite{Goertler2018}. Visual representations  of undefined uncertainty within sets have not been developed until now.

Given the scarcity of visualization methods for uncertain information in sets, we next present examples of visualization designs for each relevant orange cell from our framework from Table~\ref{tab:framework}.
\section{Design Examples for Uncertainty in Set Visualization}

It is sensible to begin the process of depicting uncertainty by constructing a visualization of the data that is 'certain', and then subsequently adapting or augmenting it as necessary to depict the uncertainty. Gershon~\cite{gershon1998visualization} calls this \emph{intrinsic} representation of uncertainty as opposed to \emph{extrinsic} representations where uncertainty information is shown in separate auxiliary displays, like a supplementary diagram or text.
The decision on whether to use intrinsic or extrinsic representations may depend on the complexity of both the data and the uncertainty. 

\subsection{Uncertain Set Membership}

\begin{figure}[h]
	\centerline{\includegraphics[width=\columnwidth]{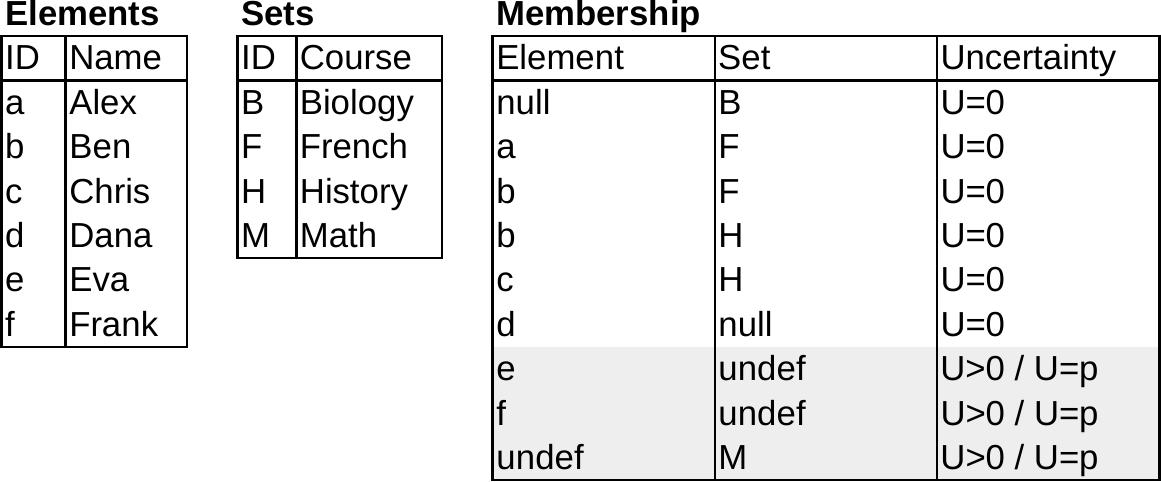}}
	\caption{Example data for sets memberships. \ccby}
	\label{fig:data}
\end{figure}

Communicating set membership is essential for set visualization. In the following, we use the toy database with students and courses from \textbf{Figure~\ref{fig:data}} for illustration. Following Cantor's~\cite{Cantor1895Sets} notation, elements are denoted by small letters, whereas sets are denoted with capital letters. For elements $a$ (Alex), $b$ (Ben), $c$ (Chris), and $d$ (Dana) membership is certain, and we also know that set $B$ (Biology) is empty. We are uncertain, however, about the membership of elements $e$ (Eva) and $f$ (Frank) as well as of set $M$ (Math). (Modeling the data in tables designed like this is just an example, more elaborate models would certainly be needed in practice).

\subsubsection{Visualizing certain set membership}

In general, certain set membership ($U=0$) can be represented in two different ways: implicitly or explicitly. Implicit representations do not use a dedicated graphical mark to visualize set membership, but rather some relation between existing marks. A common implicit example was already shown in Figure~\ref{fig:setvis}~(a) where sets are visualized as ellipses and elements of sets are visualized as dots within the ellipses. In this case, set membership is implicitly encoded through inclusion of the dots in ellipses. In addition to inclusion, also adjacency and overlap are possible implicit representations~\cite{Schulz11Implicit}.

In contrast to implicit representations, explicit representations have a dedicated graphical mark that represents set membership (in addition to marks representing sets and elements). Common explicit examples include bipartite node-link representations and matrix representations as shown in Figures~\ref{fig:setvis}~(b) and (c). In the former case, sets and elements are both depicted as dots, whereas set membership is indicated via links between set dots and element dots. The links are the explicit representation of membership. For matrix representations, sets correspond to the columns of a matrix, whereas elements correspond to matrix rows (or vice versa). The matrix cells are marked (e.g., by filling the cell) where elements are members of a set. In this case, the cells are the explicit representation of membership.

\subsubsection{Visualizing uncertain set membership}

When uncertainty needs to be considered, it is necessary to vary the representation of set membership in order to communicate either the fact that undefined uncertainty is present ($U>0$) or the exact information we may have about the defined uncertainty ($U=p$). Varying an implicit representation of set membership (i.e., inclusion, adjacency, or overlap of graphical elements) is difficult. Where in Figure~\ref{fig:setvis}~(a) should we place the dots for the uncertain elements $e$ and $f$ of our data and how should we draw the ellipse for set $M$? The problem is that graphical marks may or may not include, be adjacent, or overlap other marks, but there are no other states that could be used to indicate uncertainty. So, for implicit representations, we would first need to add further graphical marks before uncertainty could be encoded. In contrast, explicit representations already have dedicated marks for set membership, which offer several options for perceivable variation to visualize the uncertainty set memberships~\cite{MacEachren12UncertaintyVisualization}.

Next, we suggest two example designs for the case of explicit representations: (i) bipartite node-link diagrams and (ii) matrices. The uncertain set memberships (elements $e$ and $f$ and set $M$) of the data from Figure~\ref{fig:data} will be used for the purpose of illustration.

\begin{figure}
	\centerline{\includegraphics[width=\columnwidth]{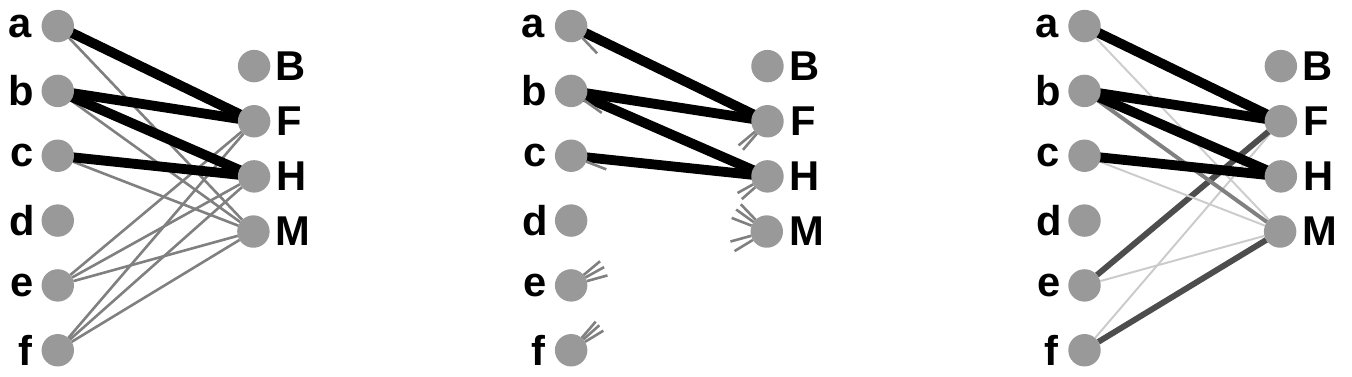}}
	\footnotesize\hspace{1em}(a) $U>0$ \hspace{6em}(b) $U>0$ \hspace{5em}(c) $U=p$
	\caption{Variants of visualizing uncertain set membership in bipartite node-link diagrams. \ccby}
	\label{fig:bipartite}
\end{figure}

\textbf{i. Bipartite node-link diagrams} In bipartite node-link diagrams, we may vary the visual properties of links to communicate uncertainty. \textbf{Figure~\ref{fig:bipartite}} shows certain set memberships as bold dark links. The figure further shows three different variants of encoding uncertainty. The fact that uncertainty is present ($U>0$) can be visualized by varying line width and color value for uncertain memberships as in (a). This makes certain memberships (bold dark lines) easily distinguishable from uncertain memberships (thin gray lines). Note, however, that elements whose membership is uncertain need to be linked to all possible sets, and vice versa for uncertain sets. This may lead to link clutter, particularly when the data has many uncertain memberships.

A design goal could thus be to reduce the resulting link clutter. Therefore, variant (b) replaces the full-length links for uncertain memberships with small link fans, which are graphically less demanding. This way, clutter can be reduced significantly, but readers need to mentally connect the elements to all possible sets.

Finally, for variant (c), we assume that we know exact probability values for possible set memberships ($U=p$). This allows us to maintain the explicit connection of elements and sets, and also to encode the different probability values per membership by varying lightness and width of lines. Thinner and lighter lines indicate lower probability values. Next, we discuss a matrix representation of set membership.

\begin{figure}
	\centerline{\includegraphics[width=\columnwidth]{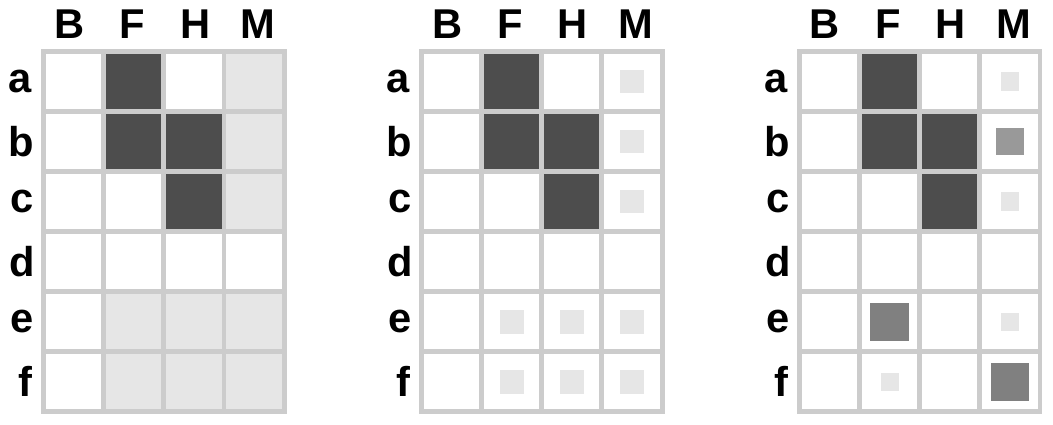}}
	\footnotesize\hspace{2em}(a) $U>0$ \hspace{5em}(b) $U>0$ \hspace{5em}(c) $U=p$
	\caption{Variants of visualizing uncertain set membership in matrix representations. \ccby}
	\label{fig:matrix}
\end{figure}

\textbf{ii. Matrices} For matrices, one can follow a similar strategy of varying the explicit representation of set membership. While we changed the graphical properties of 1D lines in the case of bipartite node-link diagrams, we now change the properties of the 2D cells of a matrix. If only the presence of uncertainty is known ($U>0$), then we can differentiate certain and uncertain set memberships by varying the fill color of matrix cells as in \textbf{Figure~\ref{fig:matrix}}~(a). However, this trivial solution might again draw too much attention to the uncertain information, simply because many cells need to be marked. To better balance the representation of certain and uncertain information, one can reduce the size of the cell marks as indicated in Figure~\ref{fig:matrix} (b).

Continuing on this line of thought, exact quantitative information about the uncertainty ($U=p$) can be encoded by varying size and color of matrix cells together as indicated in  Figure~\ref{fig:matrix}  (c). These and similar encodings in matrix cells have already been explored for visualizing multivariate graphs, in particular for showing the weight of edges~\cite{Alper13GraphComparison}.

Overall, the provided examples suggest that visualizing uncertain set memberships is rather straightforward for the cases where set membership is represented explicitly by dedicated marks. However, the visual saliency of the uncertain information and the certain information need to be cleverly balanced depending on the communicative goal of the visualization. While set membership is just a single piece of information that may be uncertain, the design of visualizations becomes more complicated when multiple uncertain set attributes need to be communicated.

\subsection{Uncertain Set Attributes}

When considering the visualization of set attributes 
the relationship between the sets and the elements is given and we are interested in visualizing an overall aggregate property of the sets - we are not interested in representing the elements themselves (indeed, there may be too many to represent explicitly). The column of our framework (Table~\ref{tab:framework}) that applies to set attributes is associated with both the other two columns: set membership relates to the attribute of set size, while other set attributes are derived from the attribute values of their elements.

For set size, there are commonly known options for the case of no uncertainty ($U=0$), and set size uncertainties ($U>0$, $U=p$) relate to uncertainty in set membership (as discussed in the previous section). This section, therefore, focuses on the more general case of set attributes derived from element characteristics. 

We use the following set scenario for illustration: There is a class of students and a list of courses that they can enroll in (e.g., Math, History, Biology, French). Students may enroll in more than one course. We consider two examples relating to attributes of the students enrolled in the classes: their residential status (domestic or international) and their age.

The two set attributes that we are interested in are the proportion of international students in each set, the 'international residential ratio', called IRR, and the average age of the students, AA. Depending on the task that the visualization is intended to support, the aggregated value (IRR or AA) may be shown for individual sub sets (e.g., intersections, differences) as well as, or instead of, for the entire sets themselves. In our example visualizations, we show both the representation of the aggregate set attributes, as well as the raw data from which they were derived.

\subsubsection{Visualizing certain set attributes}

\begin{figure}
    \footnotesize(a)
	\centerline{\includegraphics[width=0.9\columnwidth]{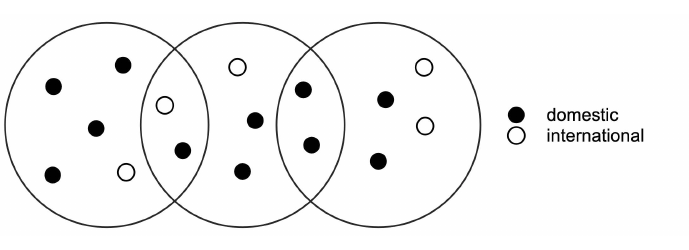}}
    \footnotesize(b)
    \centerline{\includegraphics[width=0.9\columnwidth]{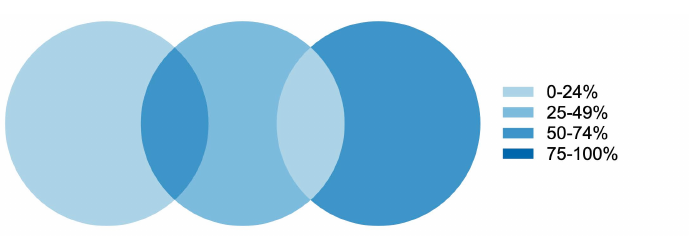}}
	\caption{(a) raw set data, showing the residential status (b) aggregate set data, showing the proportion of international students by color value. Left to right, the proportion of international students is [20\%, 50\%, 33\%, 0\%, 50\%]. \ccby}
	\label{fig:4b-certainIRR}
\end{figure}

\begin{figure}
    \footnotesize(a)
	\centerline{\includegraphics[width=0.9\columnwidth]{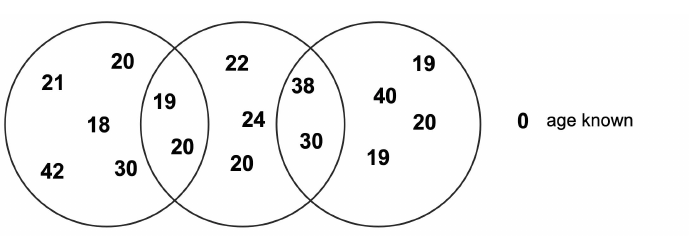}}
    \footnotesize(b)
    \centerline{\includegraphics[width=0.9\columnwidth]{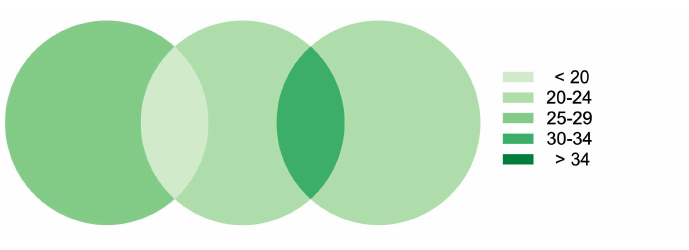}}
	\caption{(a) raw set data, showing the ages of the students (b) aggregate set data, showing average age of students by color value. Left to right, the average age is [26.2, 19.5, 22, 34, 24.5]. \ccby}
	\label{fig:4b-certainAA}
\end{figure}

The residential status and the ages of the students are the element attributes, and their enrollment in different courses is set membership. \textbf{Figure~\ref{fig:4b-certainIRR}}~(a) shows the raw data, where all elements are shown; Figure~\ref{fig:4b-certainIRR}~(b) does not show the elements, but shows a derived visualization where the IRR is shown as the desired aggregate value. For the attribute of age, \textbf{Figure~\ref{fig:4b-certainAA}}~(a) shows the age of all the students, Figure~\ref{fig:4b-certainAA}~(b) shows the AA visualization.

It is important to note that even representing set attributes without uncertainty ($U=0$) using a simple visual variable like color value has its challenges, since we cannot easily represent the aggregate value of the whole sets as well as that of the sub-sets created by the relationships between them. We focus on representing aggregate information for the sub-sets created by the relationships; additional visual variables or additional supplementary visualizations would be required if the aggregates of the entire sets are also required.

\subsubsection{Visualizing set attributes with defined uncertainty}

In the case of defined uncertainty ($U=p$), we know which courses each student is enrolled in, and we know the residential status of some students. The uncertainty lies in the fact that there are some students for whom we do not know the residential status, and/or there are some students for which we know that the information given may be incorrect. 

In this case, the visualization designer has to make choices relating to how both the aggregate value and the uncertainty are calculated. The procedure for calculating the aggregate value can:

\begin{enumerate}
	\item ignore the elements with missing values as well as those with uncertain value, or
	\item ignore the elements with missing values, and use the given values for the uncertain elements.
\end{enumerate}

The certainty can be calculated as:

\begin{enumerate}
	\setcounter{enumi}{2}
	\item the proportion of elements for which the value is certainly known, in relation to the total number of elements in the (sub) set; or
	\item the proportion of elements for which the value is certainly known, in relation to the total number of elements for which values have been given (i.e. ignoring the elements with missing values).
\end{enumerate}

\begin{figure}
    \footnotesize(a)
	\centerline{\includegraphics[width=0.9\columnwidth]{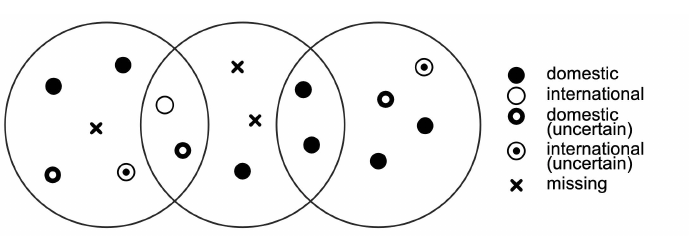}}
    \footnotesize(b)
    \centerline{\includegraphics[width=0.9\columnwidth]{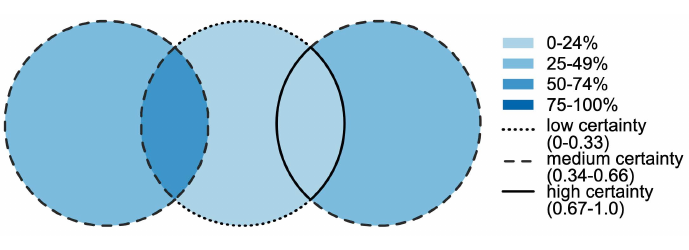}}
	\caption{(a) adaptation of the raw set data of Figure~\ref{fig:4b-certainIRR}, showing that the residential status of some students is either missing or known to be incorrect (b) aggregate set data, showing the proportion of international students represented by color value, with edges indicating the extent of uncertainty. The aggregate data calculation ignores the students with missing values, but uses the given values for all other students; the certainty calculation includes students with missing values. Left to right, the proportion of international students is [25\%, 50\%, 0\%, 0\%, 25\%] and the certainty is [0.4, 0.5, 0.33, 1.0, 0.5]. \ccby}
	\label{fig:4b-someuncertainIRR}
\end{figure}

\begin{figure}
    \footnotesize(a)
	\centerline{\includegraphics[width=0.9\columnwidth]{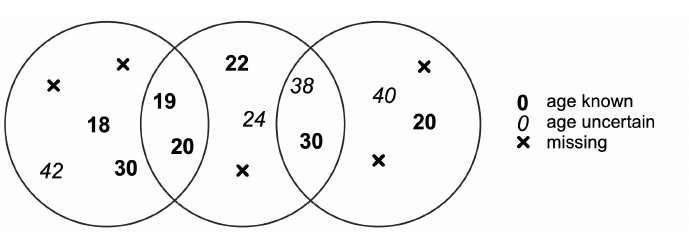}}
    \footnotesize(b)
    \centerline{\includegraphics[width=0.9\columnwidth]{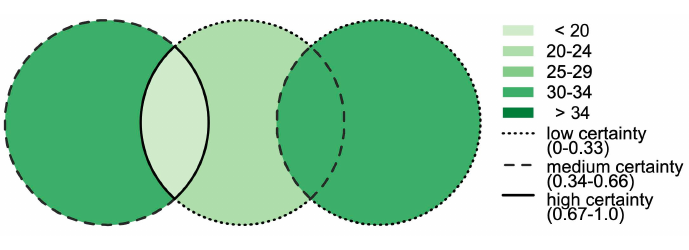}}
	\caption{(a) adaptation of the raw set data of Figure~\ref{fig:4b-certainAA}, showing that the ages of some students is unknown, and some students have given possibly incorrect ages (in italics); (b) aggregate set data, showing average age represented by color value,  with edges indicating the extent of uncertainty. The aggregate data calculation ignores the students with missing values, but uses the given values for all other students; the certainty calculation includes students with missing values.  Left to right, the average student age is [30, 19.5, 23, 34, 30] and the certainty is [0.4, 1.0, 0.33, 0.5, 0.25]. \ccby}
	\label{fig:4b-someuncertainAA}
\end{figure}

In the examples in \textbf{Figures~\ref{fig:4b-someuncertainIRR}} and \textbf{\ref{fig:4b-someuncertainAA}}, we illustrate options 2 and 3. 

\subsubsection{Visualizing set attributes with undefined uncertainty}

\begin{figure}
    \footnotesize(a)
	\centerline{\includegraphics[width=0.9\columnwidth]{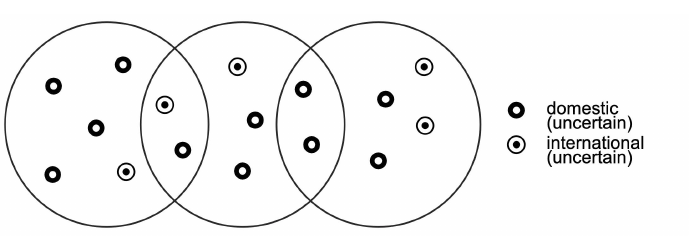}}
    \footnotesize(b)
    \centerline{\includegraphics[width=0.9\columnwidth]{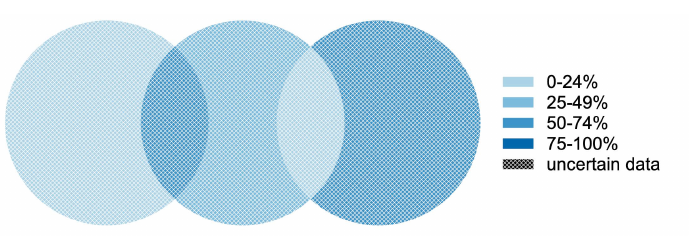}}
	\caption{(a) adaptation of the raw set data of Figure~\ref{fig:4b-certainIRR}, showing that the given residential status of all students is possibly incorrect; (b) aggregate set data, showing the proportion of international students represented by color, with texture indicating that all the data is uncertain. Left to right, the proportion of international students is the same as in Figure~\ref{fig:4b-certainIRR}: [20\%, 50\%, 33\%, 0\%, 50\%]. \ccby}
	\label{fig:4b-alluncertainIRR}
\end{figure}

\begin{figure}
    \footnotesize(a)
	\centerline{\includegraphics[width=0.9\columnwidth]{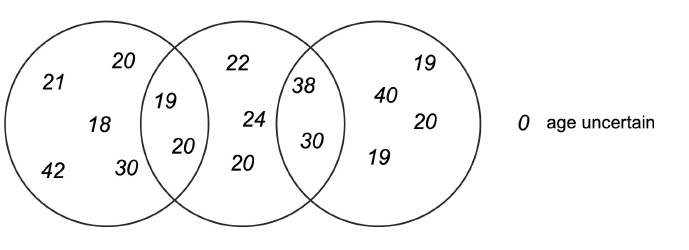}}
    \footnotesize(b)
    \centerline{\includegraphics[width=0.9\columnwidth]{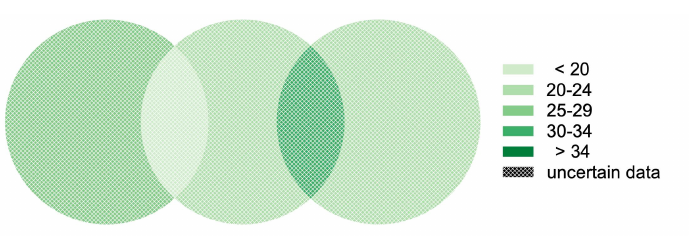}}
	\caption{(a) adaptation of the raw set data of Figure~\ref{fig:4b-certainAA}, showing that the given age of all students is possibly incorrect (in italics); (b) aggregate set data, showing the average age represented by color, with texture indicating that all the data is uncertain.  Left to right, the average age is the same as in Figure~\ref{fig:4b-certainAA}: [26.2, 19.5, 22, 34, 24.5]. \ccby}
	\label{fig:4b-alluncertainAA}
\end{figure}

In the case of undefined uncertainty ($U>0$), although we have a residential status and an age associated with each student, we know that some of this information is incorrect. Although we do not know for which students this might be the case, there is uncertainty throughout. \textbf{Figures~\ref{fig:4b-alluncertainIRR}}~(b) and \textbf{\ref{fig:4b-alluncertainAA}}~(b) show how general uncertainty can be added to the visualization using texture.  However, we recommend that when uncertainty is present throughout all the data, this should be noted as a general disclaimer statement in the caption or the text (rather than being explicitly added to the visualization by the use of additional visual variables).

\subsubsection{Possible use of other visual variables}

\begin{figure}
	\centerline{\includegraphics[width=\columnwidth]{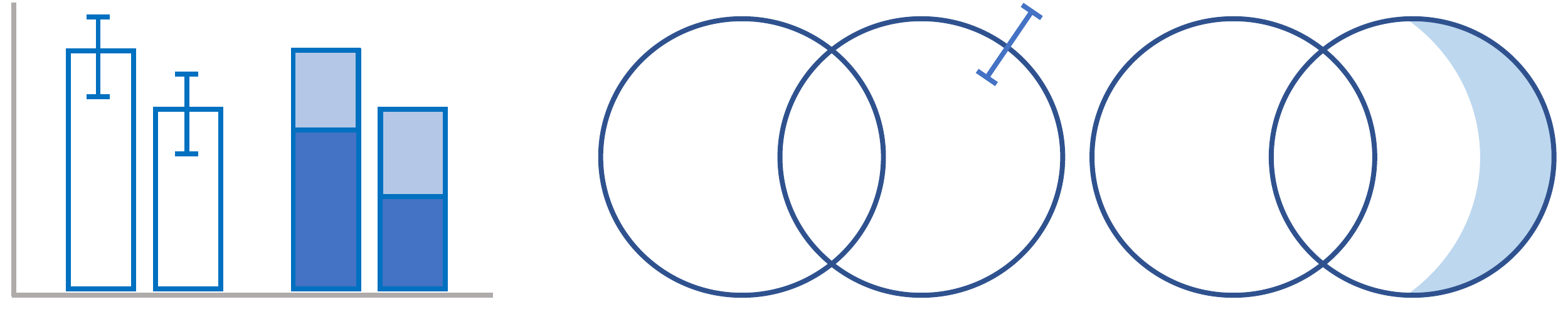}}
	\footnotesize\hspace{1em}(a) \hspace{9em}(b) \hspace{7em}(c)
	\caption{Representing uncertainty with markers indicating size variation as in (a) is not appropriate for set visualizations where the area of the set does not relate to the set attribute value as in (b) and (c). \ccby}
	\label{fig:size-markers}
\end{figure}

In the examples above, we chose to use variations in color value, line dashes, and texture to represent aggregate data values and uncertainty. While other visual variables could be used instead (for example, color hue, line weight), we claim that using size variation to represent uncertainty is inappropriate, even though it is commonly used in other data visualizations. Despite evidence that simple error bars in a bar chart are not as easy to interpret as they may appear~\cite{Correll2014}, uncertainty is often depicted with error bars or grayed out areas which indicate proportional uncertainty corresponding to the size differences as in \textbf{Figure~\ref{fig:size-markers}}~(a). It is not recommended though to apply such `size-aware’ principles to common set visualizations as in Figures~\ref{fig:size-markers}~(b) and (c), which focus on depicting set membership. The size of the graphical objects holds no meaning, and hence, inappropriate inferences could be made about the extent of the uncertainty.

\begin{figure}
	\centerline{\includegraphics[width=\columnwidth]{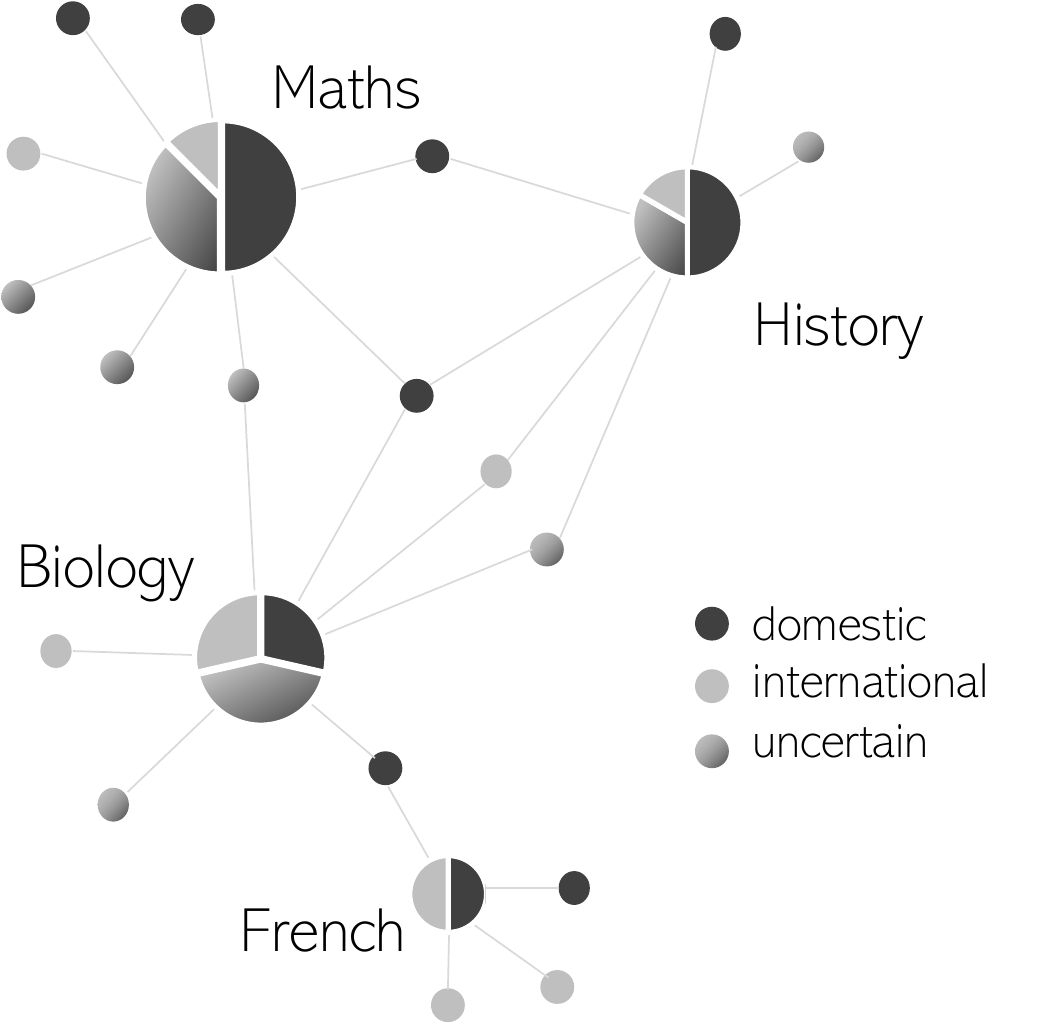}}
	\caption{Set attributes with defined uncertainty visualized in a bipartite node-link diagram. \ccby}
	\label{fig:Xa}
\end{figure}

\begin{figure}
	\centerline{\includegraphics[width=\columnwidth]{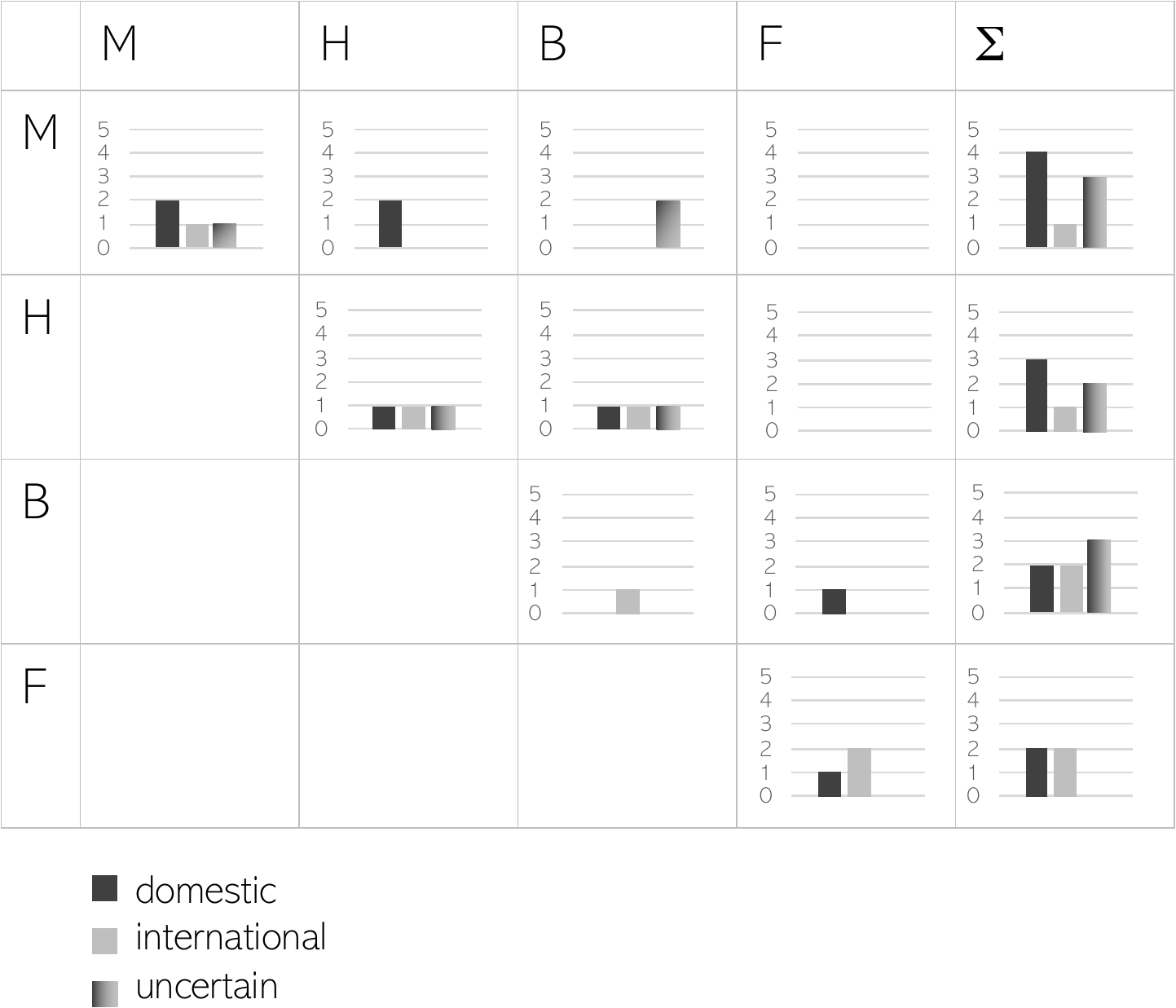}}
	\caption{Set attributes with defined uncertainty visualized in a matrix including set intersections. \ccby}
	\label{fig:Xb}
\end{figure}

\subsubsection{Adding supplementary information to depict uncertainty}

The approach above focuses on adapting an original (certain) Venn diagram visualization by adapting visual variables and existing graphical elements. An alternative approach adds supplementary graphical elements to the original diagram. For example, in a bipartite node-link diagram, sets can be represented as small pie charts indicating the proportion of international students and uncertainty (see \textbf{Figure~\ref{fig:Xa}}). In this way, any set aggregate attribute (and its uncertainty) can be added to the (certain) set representation. The association between the supplementary information and the set (or sub-set) it refers to needs to be made clear, using, for example, visual cues, like proximity or links. However, this visualization does not explicitly show (un)certainties of the intersections.
To communicate (un)certain attributes of intersections, one may use a matrix representation as in \textbf{Figure~\ref{fig:Xb}}, which shows the statistics for each intersection and the overall set.

\subsection{Uncertain Element Attributes}

Finally, we present designs for visualizing uncertain attributes of set elements. For our example, we may wish to visually represent the student age distribution enrolled in our courses. We thus focus on the age attribute of the student elements belonging to the set courses. For the defined uncertainty case ($U=p$), we know the value of an element attribute, and we also know the nature and value of the uncertainty. For example, for a given course with twenty seats, fifteen students have already enrolled and we have been given their birth year from the university registrations office. For five students also interested in enrolling in the course, we only know their age range at this point, as they ticked the age range box, i.e., 20-30 yrs. in the course enrollment questionnaire. The uncertainty lies in the fact that there are some students for whom we do not know their actual age with certainty, but we do have information on their age within a given range and above a certain threshold.

\begin{figure}
	\centerline{\includegraphics[width=\columnwidth]{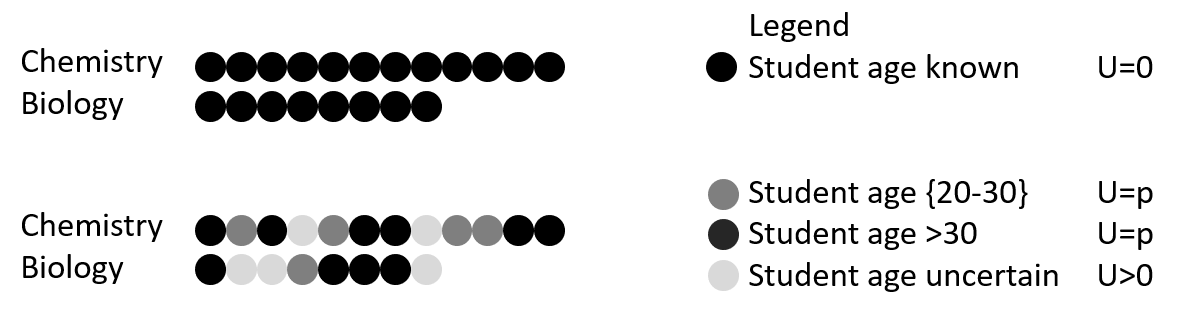}}
	\caption{Top row ($U=0$): Two courses (sets) have twenty enrolled students (set elements) where all individual ages (set element attribute) are known (i.e., black point symbols). Bottom row ($U>0$ and $U=p$): Two courses (sets) have twenty enrolled students where, a) the degree of uncertainty in student ages varies from completely unknown, that is, point symbol denoted with the lightest shade of gray, to b) mostly unknown, (i.e., age above 30 yrs.) shown with medium gray point symbols, and to c) somewhat unknown (i.e., within a given age range 20-30 yrs.), assigned dark gray point symbols). \ccby}
	\label{fig:elemattrib}
\end{figure}

Followin MacEachren et al.~\cite{MacEachren2005, MacEachren12UncertaintyVisualization}, we employ the intuitively understood visual variable color value (i.e., varying shades of gray) to denote variation of uncertainty in set element attributes, accordingly (see Figure~\ref{fig:elemattrib}). Of course there are other visual variables such as, opacity, fuzziness, texture, arrangement that we could have been used instead of color value~\cite{MacEachren12UncertaintyVisualization}. The same visual variables to show uncertainty in set elements may also be used to denote uncertainty of set membership or set attributes, as discussed before in this paper. 

In the undefined uncertainty case ($U>0$), we know the value of an element attribute, but we do not know the type and value of the uncertainty. This means for our worked example, that for a given course with twenty seats, fifteen students have already enrolled, and we have been given their birth year from the university registrations office. Five students also interested in enrolling in this course have chosen not to respond to the age class question in the enrollment questionnaire. In this case, following the logic in Figure~\ref{fig:elemattrib}, we can simply denote uncertainty with a very light gray shade, or use any other appropriate visual variable~\cite{MacEachren12UncertaintyVisualization} for those five students of whom we are uncertain about their actual age. 

For the easy case of certainty ($U=0$), we know the value of an element attribute with certainty. This means for our worked example, that for a given course with twenty seats, twenty students have already enrolled and we have been given their birth year from the university registrations office. In this case, we can depict the age attribute of our elements with any of the commonly known multivariate visualization methods~\cite{Ward15DataVis, Tominski20IVDA} and principles that are appropriate for ratio level data, i.e., dots, bar charts, box plots with commonly used visual variables~\cite{Bertin67Semiology}.

\section{Discussion}

Our investigations into how uncertainty in set data might be depicted has resulted in a novel conceptual framework representing various types of uncertainties in different aspects of set visualizations. We addressed uncertainty in set memberships, set attributes, and set element attributes, as summarized in  Table~\ref{tab:framework}. In devising and applying this new framework to data examples, we have drawn on existing visualization research. We have addressed the relevant cells in the proposed framework and also tackled challenging cells by identifying several design alternatives for set visualization designers drawing upon existing visualization research. A further outcome of our conceptual framework is the identification of challenges that warrant further detailed investigations.

\textbf{Task- and context-specific challenges.} The visualization community is well aware that user tasks and the application context are essential ingredients in designing expressive, effective, and efficient visualization solutions \cite{Schulz13Tasks, Miksch2014DesignTriangle, Tominski20IVDA}. While specific visualization tasks were considered in previous work on general set visualization~\cite{Alsallakh16SetVisSTAR}, we addressed such tasks and context only partly in our work and leave a comprehensive investigation for future research.

\textbf{Missing vs. uncertain data.} For simplicity, we treated missing and uncertain data equally in our framework. Future work should extend our proposal to disentangle these two concepts further. 

\textbf{Uncertainty propagation.} Uncertainty is not a static concept and interdependencies might occur due to data processing chains. For example, the uncertainty of a set attribute may be directly dependent on set membership and set element attribute uncertainties. Consequently, uncertainty propagation should be specifically addressed in a future version of the framework; the communication of uncertainty propagation in a set visualization will pose interesting depiction challenges.

\textbf{Temporal and spatial uncertainty.} While we considered uncertainty in set visualization, we mostly ignored the spatial and temporal context of sets, which poses additional challenges for their visualization~\cite{Fabrikant19SetSpaceTime}.
Similarly to what can be said about the dynamics of uncertainty propagation, temporal uncertainty itself also relates to time-varying changes to uncertainty, including uncertainty states with respect to points in time, perdurance, and the evolution of uncertainty in unfolding events~\cite{Gschwandtner2016}.
Likewise, uncertainty in a spatial frame of reference requires special consideration. When several domains with data uncertainty need to be understood in context, more scalable designs are needed to balance the visualization according to the needs of users~\cite{Duebel173DGeoUncertainty}.
When visualizing sets and set elements that represent spatial and temporal data, consideration will need to be given to the particular nature of spatial and temporal uncertainty.

\textbf{Perception of uncertainty depictions.} In general, the depiction of uncertainty $U$ needs to be balanced with respect to the depiction of the actual set data~$D$. Carelessly adding an uncertainty depiction to a data visualization can lead to clutter or overemphasis. For example, in our bipartite node-link diagrams from 
Figure~\ref{fig:bipartite}, uncertain set memberships require links between all uncertain set elements and all possible sets, which easily leads to visual clutter and gives much greater saliency to the uncertain information rather than to the certain information. For this specific case, we proposed small link fans, but other, more general alternatives should be explored and evaluated for their effectiveness. One such method could be reordering to reduce clutter~\cite{DBLP:journals/cgf/BehrischBRSF16}.

\textbf{Evaluation.} In general, the appropriateness, usefulness, utility, and usability of the various ways of depicting uncertainty in sets need to be established, especially as interpretations of uncertainty visualizations are dependent on the background and training of the user (e.g., graphical level, domain expertise). For example, depicting the uncertainty of a set element attribute in a Venn diagram with a lighter gray value of the element mark could be interpreted as uncertainty of set membership of that element. The evaluation of uncertainty visualizations, as a whole, is a young field with a range of important questions to be tackled.

\textbf{Design recommendations.} During the conceptualization phase of our framework, we identified design issues that lead to the following recommendations:
\begin{itemize}
\item \emph{Data first, uncertainty second}: It is easier to start with the visual encoding of the certain data, followed by the encoding of the uncertain aspects.


 \item \emph{Be aware of visual misinterpretations by the users}: Test your designs with users, as interpretation and understanding of uncertainty are likely challenging for many users; visual solutions might be misread by the target audience~\cite{Korporaal19, Kuebler20}. Ample labeling and adding legends and explanations accompanying the uncertainty visualization will help to guide the users. In some cases, we even found, it may be more effective to communicate uncertain information by non-visual means.
\end{itemize}


\section{Conclusion}

We set out to devise a conceptual framework on how uncertainty in set data could be visualized by first finding answers to still open research questions: (i) Which aspects of set type data could be affected by uncertainty, and (ii) Which characteristics of uncertainty could influence the visualization design. Based on this new conceptual framework, we then systematically discuss set visualization examples with integrated uncertainty information. We also provide a set of open challenges in the hope that these may inspire future research on uncertainty in set visualization.

\section{ACKNOWLEDGMENT}

This work was initiated by Dagstuhl Seminar 22462 on \href{https://www.dagstuhl.de/22462}{Set Visualization and Uncertainty}.



\bibliographystyle{abbrv-doi-hyperref}
\bibliography{setvisuncertainty.bib}

\end{document}